\documentclass{aa}

\usepackage{graphicx}
\usepackage{siunitx}
\usepackage[figuresright]{rotating}
\usepackage{booktabs}
\usepackage{multirow}

\usepackage{txfonts}
\usepackage[dvipsnames]{xcolor} % REMOVE LATER
\usepackage{color}
\usepackage[colorlinks=true, linkcolor=blue, citecolor=blue, urlcolor=blue]{hyperref}
\definecolor{mhi}{rgb}{0.6,0.0,0.6}

\begin{document}

\newcommand{\missingref}{\textcolor{red}{(Reference)}}

   \title{A model for the ongoing subcluster interaction in Centaurus}

   \author{Felipe S. Lohmann
          \inst{1}
          \and
          Magda Arnaboldi
          \inst{1}
          \and
          Michael Hilker
          \inst{1}
          \and
          Andreas Burkert
          \inst{2}
          \and
          Marilena Spavone
          \inst{3}
          \and
          Ortwin Gerhard
          \inst{4}
          \and
          Marina Rejkuba
          \inst{1}
          \and
          Steffen Mieske
          \inst{5}
          \and
          Tom Richtler
          \inst{6}
          \and
          Marco Mirabile
          \inst{1, 7, 8}
          \and
          Michele Cantiello
          \inst{7}
          \and
          Enrichetta Iodice
          \inst{3}
          \and
          Sara Federle
          \inst{9}
}

   \institute{European Southern Observatory, Karl-Schwarzschild-Straße 2, 85748, Garching, Germany\\
              \email{fschmidt@eso.org}
    \and
    Universitätssternwarte der Ludwig-Maximilians Universität, Scheinerstraße 1, 81679 München, Germany
    \and
    INAF -- Astronomical Observatory of Capodimonte, Salita Moiariello 16, I-80131, Naples, Italy
    \and
    Max-Planck-Institut für Extraterrestrische Physik, Giessenbachstraße, 85748 Garching, Germany 
    \and
    European Southern Observatory, Alonso de Córdova 3107, Vitacura, Santiago, Chile
    \and
    Departamento de Astronomía, Universidad de Concepción, Casilla 160-C, Concepción, Chile
    \and
    INAF -- Astronomical Observatory of Abruzzo, Via Maggini, 64100 Teramo, Italy
    \and
    Gran Sasso Science Institute, viale Francesco Crispi 7, I-67100 L'Aquila, Italy
    \and
    Instituto de Astrofísica, Departamento de Física y Astronomía, Facultad de Ciencias Exactas, Universidad Andrés Bello. Av. Fernández-Concha 700, Las Condes, Santiago, Chile.
    }

   \date{Received MM DD, YY; accepted MM DD, YY}

\date{Received 000; accepted YYY}

\abstract {Intracluster globular clusters (GCs) preserve the hierarchical and chemical history of the assembly of galaxy clusters. The nearby Centaurus cluster is believed to be undergoing an interaction between two subclusters, Cen\,30 and Cen\,45, a process that should leave an imprint on the properties of its intracluster GCs.}
{We aim to characterise the GC population in the Centaurus cluster, constraining its recent assembly history. Ultimately, we aim to construct a scenario for the geometry of the ongoing merger that is capable of explaining all the observational evidence for the interaction.}
{Using deep VLT/FORS imaging in the $V$ and $I$ bands, we select GC candidates around the centre of the Centaurus cluster based on morphology cuts and colour criteria. 
We use a Voronoi tessellation approach to study the GC spatial distribution and density profiles, demonstrating the advantages of this method over traditional annuli-based alternatives. 
We then apply a method of constraining the evolution of the galaxy luminosity function using the GC specific frequency and colour distribution, leveraging the Voronoi description of the GC spatial distribution.}
{The GC spatial distribution shows significant substructure. 
As in other clusters, red GCs are concentrated on the brightest galaxies, whereas blue GCs trace a more extended structure.
We confirm and quantify the previously reported 'bridge' of GCs which connects the two brightest cluster galaxies NGC\,4696 and NGC\,4709, and show that this structure hosts twice as many GCs as the expected number from the projection of the GC systems of the two galaxies alone.
The constraint on the evolution of the galaxy luminosity function is in agreement with the results obtained in a previous work for the Hydra\,I cluster, with a faint-end slope of $\alpha=-1.87_{-0.06}^{+0.06}$ for the galaxy population that assembled the core of the Centaurus cluster.}
{Several observational pieces of evidence point to an ongoing interaction between the two subclusters Cen\,30 and Cen\,45. We propose a nearly head-on collision scenario whereby Cen\,45 passed through Cen\,30 in the recent past, with the two subclusters moving approximately along the line of sight.
This scenario is able to explain the large line-of-sight velocity difference between the two subclusters, the hot X-ray-emitting gas in the region between them, the bridge of GCs connecting them, and the puffed-up GC system of NGC\,4709.}


   \maketitle

%-------------------------------------------------------------------

\section{Introduction}

    The hierarchical assembly process of galaxy clusters produces a stellar component that is not bound to any individual galaxy, but rather to the cluster potential as a whole. This component consists of individual stars, known as intracluster light (ICL, \citealt{montesFaintLightGroups2022}), as well as intracluster globular clusters (GCs, \citealt{whiteGlobularClustersBelonging1987,hilkerCentralRegionFornax1999}). 
    Since the ICL and ICGCs originated in the galaxies that assembled the cluster, their properties can be used to constrain the dynamical history of the cluster and the evolution of its galaxies \citep{montenegro-tabordaStellarMassComposition2025}. 
    The ICL is particularly difficult to observe due to its low surface brightness ($\mu_g \gtrsim 26$ mag/arcsec$^{2}$) and the contamination from the light of individual galaxies. In contrast, GCs are more easily detectable as they are observed as point sources which can be as bright as $g\sim$ \qty{20}{mag} \citep{chaturvediFornaxClusterVLT2022}. 
    Thus, GCs can serve as excellent tracers of the ICL, as they are typically found in the same regions of space and share similar kinematic properties (e.g. \citealt{arnaboldiKinematicsPlanetaryNebulae1996,schuberthIntraclusterGlobularClusters2008,chaturvediFornaxClusterVLT2022}). 
    Still, the study of these faint components is restricted to the nearby universe, motivating our work in the nearby Centaurus cluster ($\sim$ \qty{35.6}{Mpc}, \citealt{tullyCosmicflows32016}).

    The work by \cite{Lucey86} laid the groundwork for the study of the Centaurus cluster.
    They confirmed that the cluster is composed of two subclusters with distinct line-of-sight velocities.
    The dominant subcluster, denoted Cen\,30, recedes at \qty{3041}{\kilo\meter\per\second} with a velocity dispersion of \qty{586}{\kilo\meter\per\second} and has 2.5 times more galaxies than the smaller one, Cen\,45, which is moving at \qty{4570}{\kilo\meter\per\second} with dispersion of \qty{262}{\kilo\meter\per\second}.
    Cen\,30 and Cen\,45 host the two brightest cluster galaxies (BCGs), NGC\,4696 and NGC\,4709, respectively. Their properties are summarised in Tab. \ref{tab:galaxies}.
    Despite the large velocity difference, \cite{Lucey86} argue that the two subclusters are within the same cluster as suggested by their $U-V$ colour-magnitude diagrams, galaxy size distributions and luminosity functions.
    They favour a merging scenario, where Cen\,45 is falling into Cen\,30 along a direction close to the line-of-sight, which would explain the \qty{1500}{\kilo\meter\per\second} velocity offset.
    They note, however, that some Cen\,45 members probably belong to a background group of galaxies located in the Hubble flow.
    A similar case is found in the Virgo Cluster, where M86 is thought to be falling into M87 at a comparable relative velocity \citep{binggeliVirgoClusterHome1999}.

    Centaurus has been extensively studied in X-rays using a variety of instruments. 
    Using ASCA and ROSAT data, \cite{Churazov99} showed that the X-ray emission in the cluster is not symmetric, but rather shows an excess emission centred on NGC\,4709 and exhibits a filamentary structure. 
    They argue that this temperature excess is likely the result of gas heating due to the infall of Cen\,45 into Cen\,30. 
    This excess temperature and filamentary structure was later confirmed with XMM-Newton \citep{Walker13} and eROSITA \citep{Veronica25}, who also showed an asymmetric distribution of metals around NGC\,4696, supporting their sloshing scenario for this galaxy. 
    Finally, recent XRISM observations confirm the gas sloshing in the core of Cen\,30 happening along the line-of-sight \citep{xrismcollaborationBulkMotionGas2025}.

    Additional evidence for an ongoing interaction between the two subclusters is provided by the spatial distribution of GCs in the Centaurus cluster, studied in \cite{Federle24} and \cite{Federle25} using wide-field imaging with the Magellan telescope. In \cite{Federle24}, the authors reported that the GC system of NGC\,4696 is not azimuthally symmetric, but rather shows an excess of GCs in the direction of NGC\,4709 and NGC\,4696B. Similarly, in \cite{Federle25} they show that the GC system of NGC\,4709 is also asymmetric, with an excess of GCs in the direction of NGC\,4696, further supporting the scenario of subcluster interaction. 
    More interestingly, in the same work, they identify the existence of a 'bridge' of GCs that connects NGC\,4696 and NGC\,4709, at least in projection.
    Such a connecting structure has been reported in other galaxy clusters \citep{bassinoGlobularClusterSystems2006, dabruscoExtendedSpatialDistribution2016,ennisImprintsInteractionProcesses2024} and may be indicative of an ongoing interaction \citep{chaturvediFornaxClusterVLT2022}.

    Although the interaction scenario between the two subclusters is generally accepted, their exact geometry is still debated.
    The two BCGs in Centaurus, NGC\,4696 and NGC\,4709, lie approximately \qty{15}{arcmin} apart in projection on the sky, which corresponds to a projected separation of around \qty{155}{kpc}.
    However, the individual distances to the two galaxies are still debated, making the exact geometry of the cluster uncertain.
    Previous works estimated the distances to the two galaxies using surface brightness fluctuations (SBF) and the GC luminosity function (GCLF) as distance indicators, and concluded that NGC\,4709 is closer than NGC\,4696 by several Mpc \citep{mieskeDistanceCentaurusCluster2003,mieskeDistanceHydraCentaurus2005,Federle25}.
    In particular, \cite{Federle25} proposes that the reported GC bridge connects the two galaxies and extends for $\sim$\qty{8.5}{Mpc}.
    Other works, however, place NGC\,4709 farther away than NGC\,4696 using the $D_n$--$\sigma$, fundamental plane and Tully-Fisher methods \citep{faberSpectroscopyPhotometryElliptical1989, blakesleeSynthesisDataFundamental2001,theureauKinematicsLocalUniverse2007}.
    The most plausible scenario is that the two galaxies lie at a similar distance, as suggested by independent SBF measurements, $D_n$--$\sigma$ and statistical estimates \citep{tonrySBFSurveyGalaxy2001, willickHomogeneousVelocityDistanceData1997, tullyCosmicflows2Data2013a}, although the uncertainties in these measurements are too large for precise relative distance determination.

    The aims of this work are twofold. First, we investigate the GC population in the cluster using the deepest imaging data available to date. The substructure in the GC spatial distribution is used to shed light on the ongoing dynamical activity present in the cluster. We give special attention to the bridge of GCs connecting the two brightest galaxies, NGC\,4696 and NGC\,4709, and propose a scenario for its origin resulting from the subcluster merger between Cen\,30 and Cen\,45.

    Our second goal is to constrain the evolution of the galaxy luminosity function (LF) in Centaurus using the technique developed in \cite{lohmannIntraclusterGlobularClusters2026}, hereafter L26, which was applied to the Hydra\,I cluster. In L26, we introduced a method based on the GC specific frequency and colour distributions that constrains the number of galaxies that participated in the assembly of the cluster.
    We showed that this method is capable of recovering a faint-end slope of $\alpha\sim-1.8$ for the galaxy population that assembled the cluster, which is in agreement with the LF of high-redshift clusters (\citealt{mcleodEvolutionGalaxyStellarmass2021,navarro-carreraConstraintsFaintEnd2024}), and approaches the theoretical predictions from $\Lambda$CDM (\citealt{pressFormationGalaxiesClusters1974a,jenkinsMassFunctionDark2001}).

    The paper is organised as follows. In Sec. \ref{sec:data}, we present our data and in Sec. \ref{sec:gc_selection} the details of the GC selection.
    Sections \ref{sec:lsb} and \ref{sec:spatial_dist} present the low-surface-brightness features in our images and the spatial distribution of GCs, respectively, followed by the constraint to the evolution of the LF obtained from the GC population in Sec. \ref{sec:LF_evo}.
    Finally, we propose a model for the subcluster interaction in Sec. \ref{sec:model} and summarise our results in Sec. \ref{sec:summary}.
    
%--------------------------------------------------------------------
\section{Data}
\label{sec:data}

\begin{figure}
    \centering
    \includegraphics[width=\columnwidth]{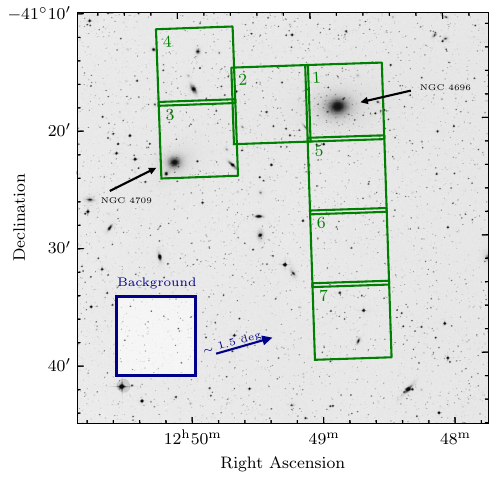}
    \caption{Footprint of our FORS observations overlaid in a DSS optical image. Green squares show the seven fields observed in the $V$ and $I$ bands. The most massive galaxies of Cen\,30 and Cen\,45, NGC\,4696 and NGC\,4709 respectively, are indicated. The background field, shown in the inset, is located $\sim$ \qty{1.5}{\degree} west of NGC\,4696.}
    \label{fig:photometry_pointings}
\end{figure}

We use deep imaging data acquired with the Very Large Telescope (VLT) equipped with the FORS1 instrument at UT1 (Observing Programme 67.A-0358(A), PI: M. Hilker) in imaging mode.
The observations were carried out in the $V$ and $I$ bands with a maximum seeing of $\leq$ \qty{0.6}{arcsec} to ensure deep photometry of point sources. 
The $V$ and $I$ observations consist of seven fields of $6.8' \times 6.8'$, covering the two BCGs, NGC\,4696 and NGC\,4709, as well as a $\sim$ \qty{20}{arcmin} strip south of NGC\,4696. The footprint of our observations is shown in Fig. \ref{fig:photometry_pointings} and the observation conditions are collected in Tab. \ref{tab:observations}.

For every field, we obtained 4 $\times$ \qty{373}{s} exposures in the $V$ band and 9 $\times$ \qty{325}{s} exposures in the $I$ band. In addition, a background frame with matching exposure times in each band was collected in an empty region located about $\sim$ \qty{1.5}{degrees} ($\sim$ \qty{0.9}{Mpc}) from the cluster centre. This allows for a statistical subtraction of background contaminants from our GC sample \citep{cantielloVEGASSSSVSTEarlytype2015,mirabileVEGASSSSTracingGlobular2024}.

\begin{table*}
\centering
\caption{Summary of our FORS observations in the $V$ and $I$ bands.}
\label{tab:observations}
\begin{tabular}{c l c c c c}
\toprule
 & Field & \# Exposures & Total Exposure Time & Airmass & Seeing \\
 &       &             & (s)           &         & (arcsec) \\
\midrule
\multirow{8}{*}{$V$}
 & Field 1 & $4$ & $1492$ & $1.11$ & $0.44$ \\
 & Field 2 & $4$ & $1492$ & $1.09$ & $0.48$ \\
 & Field 3 & $4$ & $1492$ & $1.14$ & $0.49$ \\
 & Field 4 & $4$ & $1492$ & $1.14$ & $0.59$ \\
 & Field 5 & $4$ & $1492$ & $1.08$ & $0.42$ \\
 & Field 6 & $4$ & $1492$ & $1.08$ & $0.49$ \\
 & Field 7 & $4$ & $1492$ & $1.11$ & $0.45$ \\
 & Background & $4$ & $1492$ & $1.05$ & $0.59$ \\
\midrule
\multirow{8}{*}{$I$}
 & Field 1 & $9$ & $2925$ & $1.07$ & $0.44$ \\
 & Field 2 & $9$ & $2925$ & $1.05$ & $0.48$ \\
 & Field 3 & $9$ & $2925$ & $1.25$ & $0.50$ \\
 & Field 4 & $9$ & $2925$ & $1.25$ & $0.68$ \\
 & Field 5 & $9$ & $2925$ & $1.14$ & $0.36$ \\
 & Field 6 & $9$ & $2925$ & $1.14$ & $0.42$ \\
 & Field 7 & $9$ & $2925$ & $1.07$ & $0.50$ \\
 & Background & $9$ & $2925$ & $1.10$ & $0.44$ \\
\bottomrule
\end{tabular}
\end{table*}

The data reduction steps are detailed in \cite{mieskeDistanceCentaurusCluster2003} and include bias subtraction, flat-fielding, image combining, cosmic ray removal and standard star calibration.
From the combined images, we subtract the brightest galaxies and stars with ellipse models constructed using the \textsc{photutils} Python package \citep{larrybradleyAstropyPhotutils2202025}.
We first mask bright galaxies and stars to create a 2D background map which is subtracted from the images before the galaxy and star modelling.
Then, each galaxy and star is modelled independently with an ellipse model, and the final combined model is subtracted from the image. 
% Finally, we add the median value of the 2D background map to the subtracted images to ensure that the photometry is performed on images without negative values.

On the subtracted images, we perform the source detection and PSF photometry. For this, we used the \textsc{daophot} package in IRAF \citep{todyIRAFDataReduction1986}. Standard routines\footnote{Namely: \textsc{daofind}, \textsc{phot}, \textsc{psf}, \textsc{nstar}, \textsc{allstar}.} were used for the source detection and aperture photometry, followed by PSF modeling and photometry.
The small overlaps between adjacent fields were used to cross-calibrate sources that were detected in both fields.
The analytical PSF model was constructed from a list of $\sim 20$ bright, isolated stars assuming a \cite{moffatTheoreticalInvestigationFocal1969} distribution with $\beta = 2.5$ and the sky was estimated locally within a ring around the source.
Finally, we subtract foreground galactic extinction for Centaurus based on the reddening measurements from \cite{schlaflyMeasuringReddeningSloan2011}, and correspond to $A_V = 0.312$ and $A_I = 0.171$ for the $V$ and $I$ bands, respectively. Across our fields, the reddening is approximately constant, with variations typically smaller than \qty{0.02}{mag}.
We assume the cluster distance modulus as that of NGC\,4696 ($m-M=32.75$ mag, \citealt{tonrySBFSurveyGalaxy2001}), which corresponds to a distance of 35.5 Mpc and an angular scale of \qty{10.3}{kpc/arcmin}.

\begin{table*}[t]
\centering
\caption{Properties of Centaurus' galaxies obtained from the NED and SIMBAD databases. The columns show the object name, right ascension, declination, effective radius in the $B$ band, line-of-sight velocity, $V$-band magnitude, stellar velocity dispersion, position angle and eccentricity. The values for the velocity dispersion are from \cite{hoCarnegieIrvineGalaxySurvey2011}.}
\label{tab:galaxies}

\begin{tabular*}{\textwidth}{@{\extracolsep{\fill}}lccccccccc@{}}
\hline
Object & $\alpha_{J2000}$ & $\delta_{J2000}$ & $R_e$ & $v_{\textrm{LOS}}$ & $M_V$ & $\sigma_\star$ & PA & $e$ \\

& (J2000) & (J2000) & (arcsec) & (\qty{}{\kilo\meter\per\second}) & (mag) & (\qty{}{\kilo\meter\per\second}) & (\qty{}{\degree}) & & \\
\hline

NGC\,4696 (Cen\,30)& 12:48:49.26 & -41:18:39.12 & 102.08 & $2904\pm40$ & $-22.70 \pm 0.01$ & $254\pm7.3$ & 91 & 0.28 \\

NGC\,4709 (Cen\,45)& 12:50:03.91 & -41:22:55.13 & 39.3 & $4639\pm28$ & $-21.65 \pm 0.02$ & $247\pm5.2$ & 115 & 0.17 \\

\hline
\end{tabular*}
\end{table*}

%--------------------------------------------------------------------

\section{Globular cluster selection}
\label{sec:gc_selection}

At the distance of the Centaurus cluster, GCs are unresolved point sources, with $0.6''$ seeing corresponding to $\sim$\qty{10}{pc} at \qty{35.5}{Mpc}.
The methods for selecting GC candidates follow the ones presented in L26.
First, we select point sources through morphology cuts.
In particular, we use the sharpness and $\chi$ parameters obtained from the PSF photometry.
We impose values for sharpness close to zero and a low value of $\chi$.
Finally, we restrict our point sources to a photometric error smaller than \qty{0.2}{mag}.
Appendix \ref{app:point_source_selection} describes the selection procedure in more detail and Fig. \ref{fig:point_source_selection} shows the adopted cuts in these parameters.
This selection results in a sample of 27021 point sources.

\begin{figure}
    \centering
    \includegraphics[width=\columnwidth]{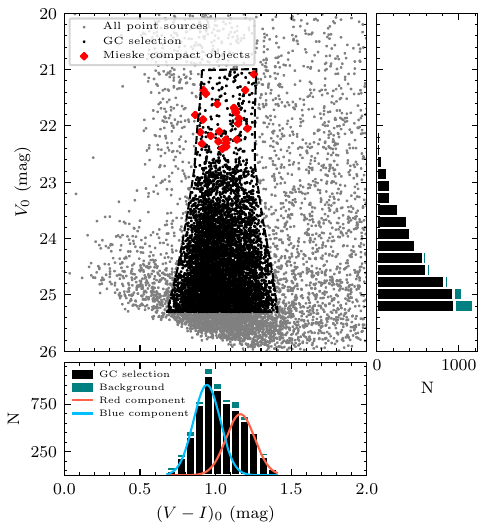}
    \caption{$VI$ colour-magnitude diagram used for the selection of GC candidates. The red diamonds are the spectroscopically confirmed ultra-compact dwarf galaxies from \cite{mieskeSearchUltracompactDwarf2007} that are used to determine the expected colour range of GC candidates. Black points show the selected candidates. The right panel shows the $V$-band luminosity function of the selected candidates and of the objects in the background field. The latter is normalised to the area covered by the seven FORS fields. The bottom panel shows the $(V-I)_0$ colour distribution of the selected candidates. The two components of the Gaussian mixture model fitted to the colour distribution are shown in the bottom panel.}
    \label{fig:cmd}
\end{figure}

Globular cluster candidates form an overdensity in the $VI$ colour-magnitude diagram (CMD).
Figure \ref{fig:cmd} shows the distribution of our point sources in the CMD.
The red diamonds show the spectroscopically confirmed sample of ultra-compact dwarfs (UCDs) from \cite{mieskeSearchUltracompactDwarf2007} which are used to select GCs, since both objects are dense stellar systems and share similar colours.
We select GC candidates down to a $V$-band magnitude of \qty{25.3}{mag}, as shown by the dashed region, yielding a sample size of 6055 candidates.
The same point-source selection process was applied to the background field, and the same region of the CMD used for GC selection was used to estimate the number of contaminants, resulting in 90 objects.

The photometric completeness in each field is evaluated by performing injection-recovery experiments of artificial stars.
We use the IRAF package \textsc{artdata} to inject 100 artificial stars with a uniform distribution in space and magnitude into our images, and then perform the PSF photometry in the injected images with the same parameters used in the point source selection to evaluate the recovered fraction.
This process is repeated 100 times with different shufflings of the artificial-star magnitudes and positions.
Figure \ref{fig:completeness} shows the completeness curves for all fields, in the $V$ and $I$ bands.
We note that field 4 is significantly less complete than the others due to worse seeing conditions, with a completeness fraction of $\sim$50\% at our GC magnitude limit of $V=23.5$ mag.

We use the completeness curves to derive the total number of GCs in each field. First, we determine a limiting $V$-band magnitude where the completeness is at least 95\% in both bands.
We consider a blue colour limit of $(V-I)_0=$ \qty{0.8}{mag} such that the limiting magnitude is $V_\textrm{lim} = \textrm{min}(V_\textrm{95}, I_\textrm{95} + 0.8)$.
Then, we analyse the GC luminosity function up to $V_\textrm{lim}$ and subtract the number of background objects in each magnitude bin. Note that the background field has a higher completeness than the cluster fields, and thus is always complete at $V_\textrm{lim}$.
Finally, we fit it with a Gaussian function centred in $M_V=$ \qty{-7.5}{mag} \citep{rejkubaGlobularClusterLuminosity2012} and with a $\sigma$ free to vary between 0.80 and \qty{1.35}{mag} \citep{Villegas10}.
The total number of GCs expected inside the field corresponds to the integral of the fitted Gaussian divided by the magnitude bin size that was used for the fitted luminosity function.
In Fig. \ref{fig:GCLF_fits}, we show the Gaussian fit to the GCLF in each field and the predicted total number of GCs in them.

The GC colour distribution in early-type galaxies is typically bimodal. This feature is usually interpreted as a result of two distinct populations of GCs: one blue, metal-poor and one red, metal-rich \citep{muratovModelingMetallicityDistribution2010}.
This bimodality can also result from the nonlinearity of the metallicity-colour relation, and can be present even in unimodal metallicity distributions \citep{yoonExplainingColorDistributions2006, cantielloMetallicityColorRelationsBimodal2007}.
However, we can assume that the global GC population in a galaxy cluster is more complex and not unimodal in metallicity.
It is challenging to confidently separate our GCs into the two families, as there is a significant colour overlap between the two, but still, we determine a colour cutoff to roughly separate our GC sample into blue (bGCs) and red (rGCs).
This is done by fitting a two-component Gaussian Mixture Model to the colour distribution and obtaining the colour where the two components intersect.
The bottom panel of Fig. \ref{fig:cmd} shows the colour distribution of our sample together with the fitted model, which produces a colour cutoff of $(V-I)_0=$ \qty{1.06}{mag}. Our bGC and rGC samples contain 3611 and 2444 objects, respectively.

\subsection{Voronoi tessellation}
\label{sec:voronoi}

We study the GC spatial distribution in Centaurus by performing a Voronoi tessellation of our FORS footprint \citep{dornanUtilizingVoronoiTessellations2024a}. \cite{dornanMajorMergersMean2025} showed that this approach produces more accurate density profiles and better preserves substructures in the number density profiles compared to an annulus-based method.
In essence, space is segmented in cells containing one GC, where each cell comprises the points in the sky that are closer to the host GC than to any neighbouring GC.
In this case, the GC number density in each cell is simply the inverse of its area.
Then, neighbouring cells can be merged to form larger cells that host more GCs for a larger statistical significance in the number density.

The merging of cells follows the steps presented in \cite{dornanUtilizingVoronoiTessellations2024a}, where higher density cells are merged first.
First, all cells are sorted by decreasing density, then the first cell (highest density) is merged with its $n$ closest neighbours.
The merged cells are removed from the list, and the process repeats until the list is empty.
We note that this method still results in some cells containing only one GC, whose neighbours were merged with other cells and left them isolated.
To treat this, these cells are merged with their highest density neighbour after the first round of merging is performed, thus ensuring that every cell contains, at least, the specified minimum number of GCs.

The number density of each cell corresponds to the number of GCs inside the cell divided by its area. To correct for photometric incompleteness and contamination from fore- and background objects, we scale the number densities by a correction factor that corresponds to the ratio between the expected total number of GCs and our observed sample size.
This correction factor varies between fields due to their different completeness levels, but lies between 1.5 and 2.0 for fields 1, 2, 3 and 5. Field 4 has a higher correction factor of 4.0 due to its lower completeness, while fields 6 and 7 are dominated by background objects and thus have a factor lower than unity, at around 0.8.
An independent correction factor is obtained for rGCs and bGCs separately, and its values are shown in the legend of Fig. \ref{fig:GCLF_fits}.

\begin{figure*}
    \centering
    \includegraphics[width=\textwidth]{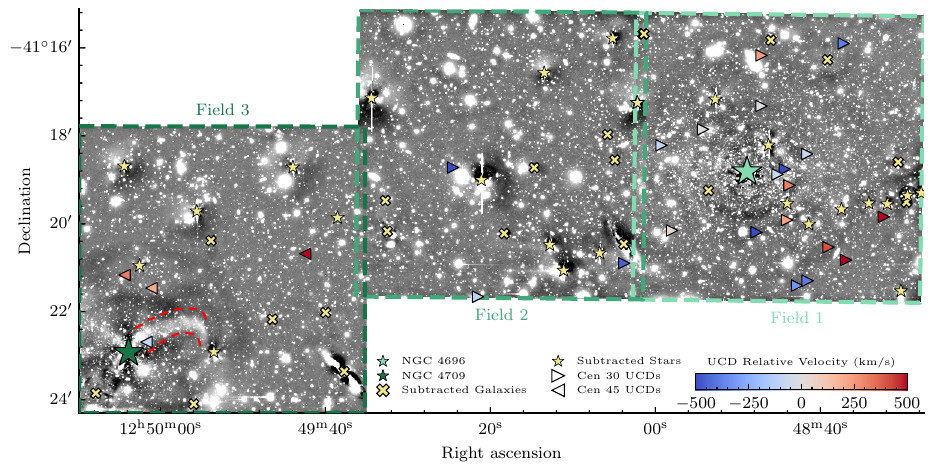}
    \caption{Galaxy- and bright-star-subtracted images of fields 1 to 3, from right to left. The two brightest galaxies, NGC\,4696 and NGC\,4709, are marked with large asterisks on fields 1 and 3, respectively. The subtracted galaxies and stars are indicated with small crosses and asterisks, respectively. The UCDs from \cite{mieskeSearchUltracompactDwarf2007} are shown as horizontal triangles, and are colour-coded according to their residual velocities with respect to their host group, Cen\,30 or Cen\,45. The red dashed line indicates the plume that emerges from NGC\,4709.}
    \label{fig:subtracted_images}
\end{figure*}

\section{Subtracted images of NGC\,4696 and NGC\,4709}
\label{sec:lsb}

In Fig. \ref{fig:subtracted_images}, we show the galaxy- and star-subtracted images of fields 1 to 3, numbered according to Fig. \ref{fig:photometry_pointings}. 
Field 1 contains the brightest galaxy in Cen\,30, NGC\,4696, while field 3 contains the brightest one in Cen\,45, NGC\,4709. 
Field 2 is located between the two galaxies and contains a mix of various galaxies and, as will be discussed, many GCs.
Inspecting the subtracted image of NGC\,4696, we detect a striking number of point sources around the inner regions of the galaxy, which are mostly GCs associated with the galaxy.
% In a radius of \qty{2}{arcmin} (\qty{20.6}{kpc}), we detect 1626 point sources, making NGC\,4696 an extremely rich GC system.
Our image also reveals the presence of faint shells around the galaxy that extend out to a distance of $\sim$ \qty{1.2}{arcmin} (\qty{12.4}{kpc}) from the galaxy centre. 
This structure suggests a major merger within Cen\,30 may have contributed to the assembly of NGC\,4696 \citep{popFormationIncidenceShell2018}.
However, NGC\,4696 exhibits a complex dust distribution, with a prominent dust lane in its core and significant substructure at larger distances as revealed by soft X-ray images (see Fig. 1 in \citealt{sandersVeryDeepChandra2016}). The dust could therefore offer an alternative non-stellar origin for the observed 'shells' around NGC\,4696.

NGC\,4709, marked with the dark green asterisk in field 3, also exhibits low-surface-brightness features around it.
Most notably, we detect a significant plume emerging from the galaxy towards the northwest, which warps downwards as it extends out.
Unlike shells, tidal streams are formed from the infall of satellites in less eccentric orbits \citep{amoriscoFeathersBifurcationsShells2015,karademirOuterStellarHalos2019}.
A less significant fan-like feature appears to the northeast of the galaxy, but is difficult to confirm due to our limited spatial coverage.  

Across all fields, we plot the positions of the spectroscopically confirmed UCDs from \cite{mieskeSearchUltracompactDwarf2007} that overlap with our observations.
We separate the subcluster membership of the UCDs with a velocity cut at \qty{4000}{\kilo\meter\per\second}.
The UCDs belonging to Cen\,30 are plotted as triangles pointing to the right, while those belonging to Cen\,45 are plotted as triangles pointing to the left. They are coloured according to the residual velocity with respect to their host group.
The velocity distribution of UCDs hints at a weak pattern, albeit with low number statistics, in particular for the Cen\,45 UCD sample.
In the region between NGC\,4696 and NGC\,4709, the Cen\,45 UCDs seem to be redshifted with respect to NGC\,4709. 
Conversely, the Cen\,30 UCDs seem to be preferentially blueshifted with respect to NGC\,4696 in the same region, while the redshifted UCDs seem to be located on the opposite western side.
This pattern may be indicative of large-scale rotation of the two subclusters, although a larger spectroscopic sample is needed to confirm this claim.

%--------------------------------------------------------------------

\section{Globular cluster spatial distribution}
\label{sec:spatial_dist}
    
    \begin{figure*}
    \centering
    \includegraphics[width=0.8\textwidth]{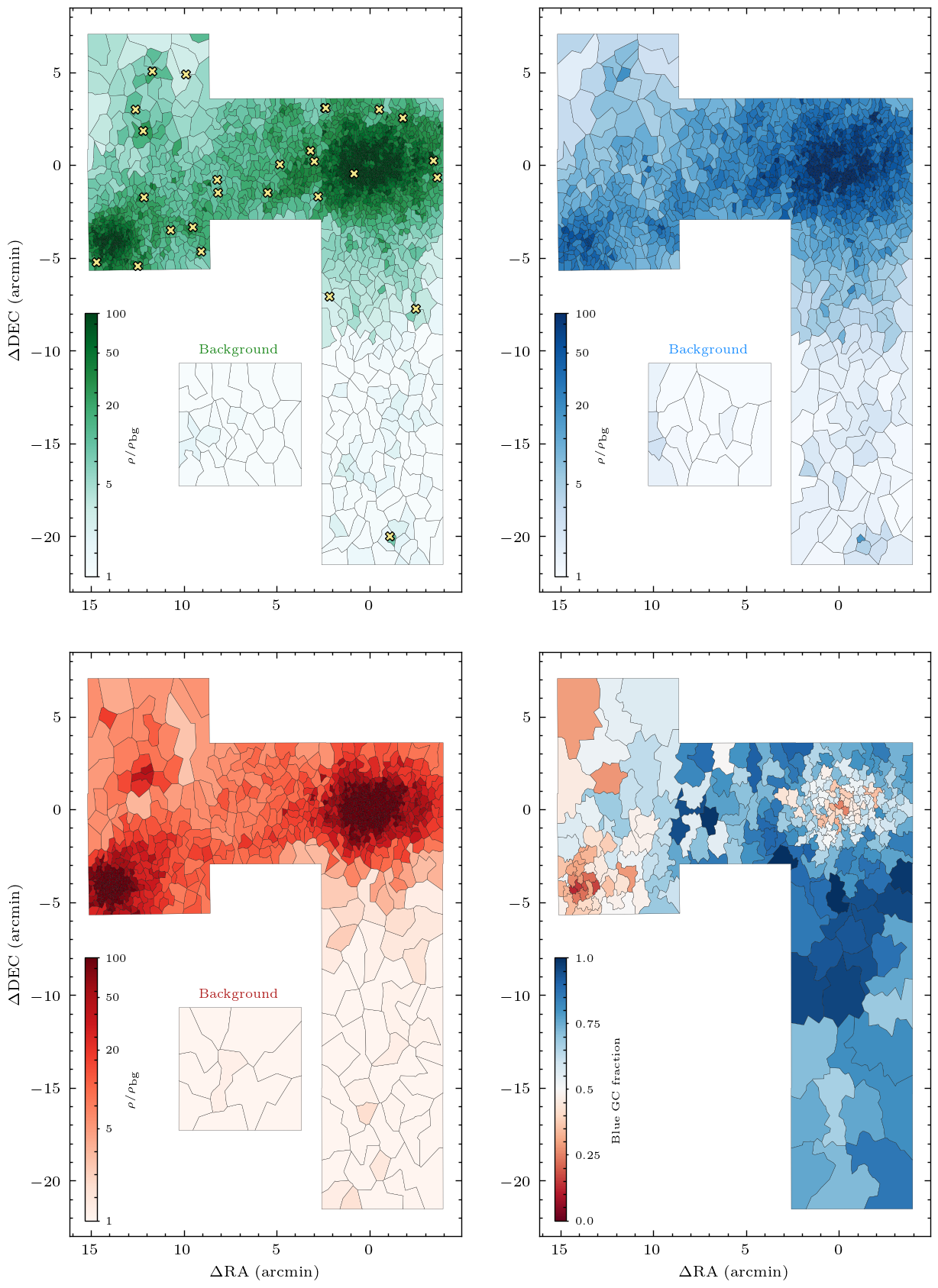}
    \caption{Spatial distribution of the Voronoi cells relative to the centre of NGC\,4696 for the full GC sample (top left), bGCs (top right) and rGCs (bottom left). The cells are coloured according to their GC density and follow steps in multiples of the respective background density. The yellow crosses in the top left panel show the positions of the subtracted galaxies, as in Fig. \ref{fig:subtracted_images}. The bottom right panel shows Voronoi diagram with larger cells, with at least 20 GCs per cell, which are coloured according to the fraction of bGCs in the cell.}
    \label{fig:spatial_density_maps}
    \end{figure*}

    Figure \ref{fig:spatial_density_maps} shows the spatial density maps obtained from the Voronoi tessellation method described in Sec. \ref{sec:voronoi}, with a minimum number of 3 GCs per cell. The cells are displayed individually and are coloured according to their completeness-corrected GC density, in units of the background density. The top left panel shows the full GC sample, while the top right and bottom left panels show the two colour subsamples.
    
\begin{figure}
    \centering
    \includegraphics[width=\columnwidth]{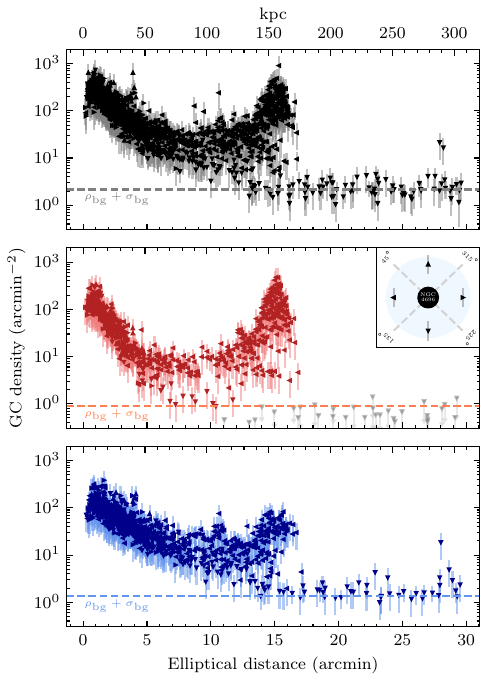}
    \caption{GC number densities from the Voronoi cells for the full GC sample (top), rGCs (middle) and bGCs (bottom), as a function of elliptical distance to NGC\,4696. The triangle markers are oriented according to the cell's position angle, following the schematic shown in the inset on the middle panel.}
    \label{fig:number_density_profiles}
    \end{figure}

    Unsurprisingly, the densest regions are centred around NGC\,4696 and NGC\,4709, with densities reaching $\sim 200$ times the background value (see Fig. \ref{fig:number_density_profiles} for exact values).
    The maps also reveal local overdensities which spatially correspond to subtracted cluster galaxies in our images.
    The most striking feature, however, is the significant overdensity of GCs connecting the two brightest galaxies.
    This bridge of GCs was also reported in \cite{Federle25}, who suggested that it originates from the interaction between the two subgroups and extends for several megaparsecs along the line of sight.
    The nature of this bridge will be discussed in Sec. \ref{sec:model}.
    Along the bridge, we also identify a significant GC overdensity emerging to the southeast of NGC\,4696, in the direction of NGC\,4709. Interestingly, this structure is surrounded by galaxies that were subtracted from the image. Therefore, it is not clear whether this overdensity is associated to these galaxies or instead represents real substructure in the intracluster GC population.

    When the spatial distributions of the red and blue GC subsamples are analysed individually, we observe important differences.
    The bGCs are spatially more extended than rGCs, which can be verified by comparing the sizes of the Voronoi cells of the two samples both along the bridge and the southern direction.
    This also means that bGCs contribute more to the bridge population than rGCs, although the overdensity is also clear in the red population.
    Around the bright galaxies, the two populations differ as well: not only are bGCs more extended, they also are distributed in a more irregular and patchy configuration compared to the tightly concentrated and boxy distribution of rGCs.
    These differences are not uncommon (e.g., \citealt{cantielloFornaxDeepSurvey2020,durrellNextGenerationVirgo2014,leeDetectionLargeScaleStructure2010,lohmannIntraclusterGlobularClusters2026}), and agree with the scenario of different origins for the two GC populations: red GCs are metal-rich and originate in massive galaxies and represent the 'in-situ' GC component, while bGCs are metal-poor and come from dwarf galaxies that are disrupted at larger distances from the cluster centre \citep{amoriscoContributionsAccretedStellar2017,hilkerCentralRegionFornax1999,ramos-almendaresIntraclusterGlobularClusters2018}.

    Another interesting feature in the distribution of bGCs can be seen around NGC\,4709.
    There, bGCs have significantly lower number densities when compared to rGCs.
    This is not the case for NGC\,4696 where both bGCs and rGCs have similar densities, a contrast that can also be clearly seen in Fig. 7 of \cite{Federle25}.
    This difference is unusual, as both galaxies have similar masses (within a factor of 3) and might be expected to share a similar population of accreted bGCs.
    As will be discussed in Sec. \ref{sec:model}, an explanation for this difference is a scenario where NGC\,4709 was stripped of its bGCs as a result of the interaction with NGC\,4696.
    
    In the bottom-right panel of Fig. \ref{fig:spatial_density_maps}, we present another Voronoi segmentation, now with a much higher number of GCs per cell of at least 20.
    We colour the cells according to the fraction of bGCs inside the cell, $N_{\textrm{bGCs}}/N_{\textrm{cell}}$.
    These numbers were corrected for incompleteness and background contamination in the same way as the number densities shown in the other panels.
    As hinted by the density maps, we confirm that rGCs dominate the inner regions of the massive galaxies, while bGCs prevail in the intracluster region.
    Additionally, we see again that, around NGC\,4709, the fraction of bGCs is significantly lower than in NGC\,4696.
    We verify that these features are independent of the binning in terms of the minimum number of GCs per cell.

    For each density map in Fig. \ref{fig:spatial_density_maps}, we plot the number density profile as a function of elliptical distance to NGC\,4696 in Fig. \ref{fig:number_density_profiles}.
    The elliptical distance of a cell is defined as the semimajor axis of the ellipse with position angle and ellipticity of NGC\,4696 (Tab. \ref{tab:galaxies}) that intersects the cell centre.
    In the figure, the number densities for all cells are plotted with triangular markers that point towards their approximate position with respect to NGC\,4696.
    In other words, values for cells along the bridge are plotted as triangles pointing left, while for those located to the south are plotted as triangles pointing down.
    The completeness-corrected number densities are plotted with Poissonian errors, namely the square root of the number of GCs inside the cell.

    It is immediately apparent how the density profiles form two separate sequences: a high density track to the east corresponding to the bridge, and a lower density one to the south that steadily falls to background levels.
    The southern sequence reaches the background density at around $\sim$ \qty{13}{arcmin}, while the bridge density remains around a dex higher at the same distance.
    Note also how local overdensities appear in these profiles as outliers compared to the majority of the cells, which form smooth sequences with low variance. 
    These substructures would be erased if the number density profile was constructed as radial averages, demonstrating the advantage of the Voronoi method.

    When we compare the red and blue subsamples, their different spatial distributions become even more apparent.
    The inner slope is significantly steeper for rGCs.
    Also, according to the southern sequence, they fall to the background density at $\sim$ \qty{8}{arcmin}, much earlier than bGCs at $\sim$ \qty{15}{arcmin}.
    In both cases, the bridge overdensity exists but seems to be dominated by bGCs that contribute with a higher number of cells in the region.
    Interestingly, the cell density along the bridge is only slightly higher for bGCs, with a difference of around \qty{0.1}{dex}.
    The fact that the two samples have a different number of cells along the bridge, but comparable densities, implies that blue and red GCs are not distributed in the same way.
    Indeed, by comparing the two distributions in Fig. \ref{fig:spatial_density_maps}, we see that rGCs along the bridge follow a narrow filament between the two galaxies, very different from the patchy distribution of bGCs that spreads out inside field 2. Part of this structure is likely due to the GC systems of the subtracted galaxies in this field (see top left panel).

\begin{figure}
    \centering
    \includegraphics[width=\columnwidth]{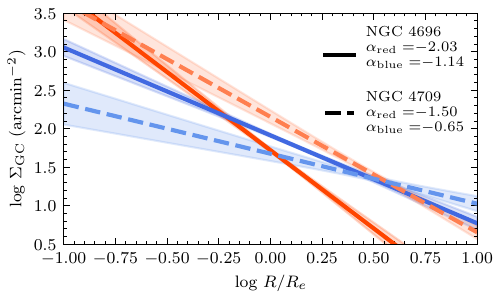}
    \caption{Linear fit to the number density profile of red and blue GCs in units of effective radii, performed on the Voronoi cells located within a projected distance of $1 \leq R \leq 5$ arcmin of the galaxy. The shaded regions show the 2$\sigma$ confidence interval of the fit obtained from bootstrapping. The best fit slopes are indicated in the legend.}
    \label{fig:GC_profiles}
    \end{figure}

    The GC spatial distribution reveals that the radial profiles of red and blue GCs differ between NGC\,4696 and NGC\,4709.
    We quantify this difference in Fig. \ref{fig:GC_profiles}, where we perform a linear fit to the values of Voronoi cells located within a projected distance of $1 \leq R \leq 5$ arcmin of the two galaxies and plot the profiles in units of galaxy effective radii (Tab. \ref{tab:galaxies}).
    Both red and blue GCs show a shallower profile in NGC\,4709 when compared to their counterparts in NGC\,4696.
    This is an important difference that indicates that the GC system of NGC\,4709 is perturbed.
    In Sec. \ref{sec:model} we will provide a model for the interaction between the galaxies that is able to account for such behaviour.

\begin{figure}
    \centering
    \includegraphics[width=\columnwidth]{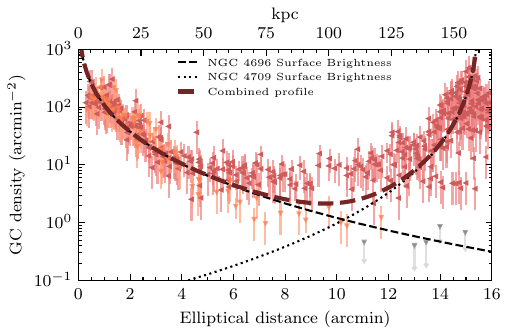}
    \caption{Red GC density profile along two directions: east of NGC\,4696, shown in red, and south of NGC\,4696, shown in coral.
    The surface brightness profiles of NGC\,4696 and NGC\,4709 are shown as dashed and dotted lines, respectively, and are vertically shifted to match the inner GC density profiles.}
    \label{fig:bridge_sersic_profiles}
    \end{figure}

    Although the GC bridge represents a real overdensity well above the background level, it is still plausible that it may originate from a projection effect.
    Namely, the observed structure could be a simple superposition of two undisturbed GC systems associated to NGC\,4696 and NGC\,4709, rather than the result of an interaction between the two galaxies.
    We explore this possibility in Fig. \ref{fig:bridge_sersic_profiles}.
    In this figure, we plot the rGC density profile along the bridge and along the southern direction, as well as the surface brightness profiles of NGC\,4696 and NGC\,4709. These profiles were obtained from wide-field imaging data from the VLT Survey Telescope (VST) as part of the VEGAS\footnote{https://sites.google.com/inaf.it/vegas/home} survey \citep{iodiceVSTEarlytypeGAlaxy2021} and will be described in detail in Spavone et al. (in prep.).
    In order to compare the GC number density with the light distribution, we transformed the surface brightness (in mag/arcsec$^2$) into surface luminosity (in $L_\odot$/arcsec$^2$) and arbitrarily shifted the resulting profiles to match the inner GC density profiles of the two galaxies.
    The $r$-band surface brightness profile from the VST was converted into the $V$ band using the relation by \cite{kostovTransformationPanSTARRS1Gri2018}, $M_V = M_r - 0.017 + 0.492(g-r)$, assuming a constant $g-r=0.7$ mag.

    Red GCs are used for this comparison as they are more tightly correlated with the galaxy light, as is commonly observed in other galaxy clusters (e.g. \citealt{schuberthGlobularClusterSystem2010, cantielloVEGASSSSIIComparing2018, lohmannIntraclusterGlobularClusters2026}).
    Encouragingly, the GC density profile south of NGC\,4696 is in excellent agreement with the galaxy's surface brightness profile.
    Since the plot is in the reference frame of NGC\,4696, it is difficult to compare the light profile of NGC\,4709 with the cells perpendicular to the bridge. Still, we find a good agreement in the inner regions of the galaxy.
    If instead we used the full GC sample for this test, the effect would be even more pronounced, as the full sample has a higher density in the bridge region than the rGC sample (see the top and middle panels of Fig. \ref{fig:number_density_profiles}).

    In the case of a projection effect, we would expect the combined profile of the two galaxies to trace the GC bridge. 
    This is clearly not the case.
    The bridge lies significantly above the combined surface luminosity profile, containing roughly twice the number of expected GCs at a distance of \qty{8}{arcmin}.
    This demonstrates that the overdensity is real and further supports the idea that it is a result of an interaction between the two subgroups Cen\,30 and Cen\,45.

\section{Reconstructing the past luminosity function of Centaurus}
\label{sec:LF_evo}

In L26, we developed a method to constrain the evolution of the galaxy LF of the Hydra\,I cluster using the GC specific frequency and colour distribution.
As a result, we derive the number of galaxies which participated in the assembly of the cluster that are necessary to reproduce these two observables. 
We summarise the method below but refer to L26 for an extensive description.

The GC specific frequency, defined as $S_N\equiv N_{GC} 10^{0.4(M_V + 15)}$, measures the number of GCs per unit luminosity of a galaxy, thus quantifying the richness of its GC system \citep{harrisGlobularClustersGalaxies1981}.
The $S_N$ has a known dependence on the galaxy luminosity, with a characteristic U-shaped relation in which it is higher for both dwarf and giant elliptical galaxies, with a minimum close to $M_V \sim -20$ mag \citep{georgievGlobularClusterSystems2010}. While the $S_N$ is typically measured for individual galaxies, it can also be used to quantify the GC richness in the intracluster space.
In L26, we used the intracluster $S_N$ together with the underlying integrated light to constrain the population of dwarf galaxies disrupted during the assembly of the cluster. We divided the intracluster region into radial bins centred on the BCG and measured the $S_N$ in each bin. We then inverted the $S_N$-$M_V$ relation from \cite{georgievGlobularClusterSystems2010} to infer the characteristic $M_V$ of a dwarf galaxy associated with the measured $S_N$.
For each radial bin, we consider a Gaussian luminosity function centred on this characteristic $M_V$ with a width of $\sigma=0.9$ mag and divide it by the typical GC occupation fraction \citep{carlstenELVESIIGlobular2022}, resulting in a skewed Gaussian LF that is interpreted as the number of galaxies that contributed to the individual radial bin (see Appendix G in L26).
Finally, we normalise this luminosity function to match the measured luminosity in the bin obtained by integrating the surface brightness profile of the BCG, including the ICL.
By combining the contribution from all radial bins, we reconstruct the faint end of the LF of the galaxies that assembled the cluster. 

In this work, we apply this methodology leveraging the spatial information preserved by the Voronoi cells. Instead of radial bins, we apply this technique on a bin-by-bin basis, where each Voronoi cell provides a differential contribution to the LF. Using the Voronoi maps with 3 GCs per cell or more, we obtain the total number of GCs in each cell from its completeness-corrected GC number density, and the integrated light from the azimuthally averaged surface brightness profile of NGC\,4696 (Spavone in prep.).
To obtain the cell luminosity, we identify the closest elliptical isophote to the cell centre and integrate its surface luminosity over the area of the cell.

The $S_N$ measured in the cell is converted into a skewed Gaussian luminosity function in the same way as described above for the annuli-based approach.
The contribution from all cells corresponds to the reconstructed faint end of the LF, which is shown as orange symbols in the left panel of Fig. \ref{fig:LF_constraint}.
To avoid contamination from the GC system of NGC\,4709, we mask a circular region of \qty{5}{arcmin} around the galaxy.
The reconstructed LF\footnote{This method contains important assumptions, as detailed in L26. The $S_N$--$M_V$ relation exhibits significant intrinsic scatter, which we account for by introducing a skewed Gaussian progenitor luminosity distribution. However, the adopted form and width of this distribution are not uniquely constrained, and alternative treatments of the scatter may yield different results. The inferred LF also depends on the GC occupation fraction. The applicability of the relation from \cite{carlstenELVESIIGlobular2022} to the Centaurus cluster is another source of uncertainty. Furthermore, we assume that the ICGCs originate from the complete disruption of dwarf galaxies, whereas some may instead have been stripped from galaxies that survive to the present day. A comprehensive quantification of the systematic uncertainties associated with these assumptions is beyond the scope of this paper.} from the Voronoi-based approach yields a Schechter slope of $\alpha=-1.87_{-0.06}^{+0.06}$, which is steeper than the present-day value of $\alpha=-1.14$ \citep{Misgeld09}.
The total number of dwarf galaxies, obtained by summing the contributions across all magnitude bins in the reconstructed LF, is approximately 4000.
The past slope is in agreement with the value reported for the Hydra\,I cluster in L26. It has smaller uncertainty, possibly due to the more accurate spatial information provided by the Voronoi tessellation compared to the annuli-based approach.
This value is also in agreement with theoretical predictions for dark matter halos \citep{pressFormationGalaxiesClusters1974a}, as well as with high-redshift observations \citep{mcleodEvolutionGalaxyStellarmass2021,navarro-carreraConstraintsFaintEnd2024}.
The isophotal ellipses of NGC\,4696 are highly elongated along the east-west direction, approximately aligned with the GC bridge.
As a result, only the Voronoi cells in this region can be used to constrain the faint end of the past luminosity function.
Still, we find a steeper Schechter slope that indicates that the east-west direction seems to be to preferred direction of the past accretion. 

The bright end of the LF is constrained using the colour distribution of GCs, which preserves information about the progenitor galaxies that merged together to form the currently observed BCG.
In the Virgo cluster, \cite{pengACSVirgoCluster2006} showed that the colour distribution of GCs differs among galaxies of different luminosities, with bright galaxies hosting a bimodal GC population and faint galaxies hosting a unimodal, metal-poor population.
In L26, we used the GC colour distribution of Virgo galaxies as a library of templates to fit the colour distribution of GCs in Hydra\,I, and thus constrain the number of galaxies of each template that likely contributed to the assembly of the BCG.
The fit was performed in circular annuli centred on the BCG, out to a distance of $\sim$ \qty{80}{kpc}.
The contribution from all bins correspond to the constrained bright end of the LF.

\begin{figure*}
\centering
\includegraphics[width=\textwidth]{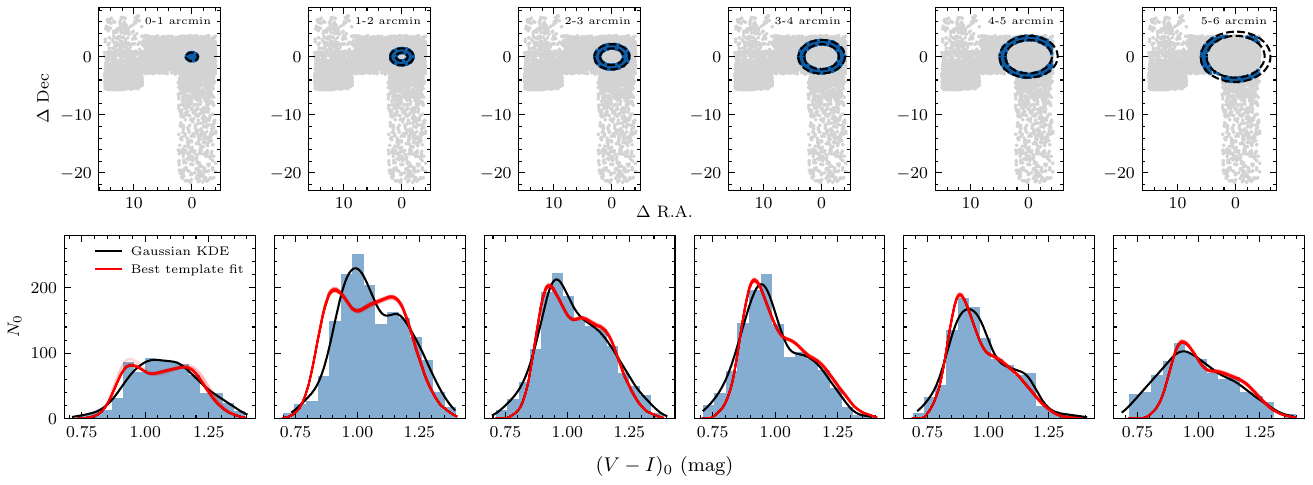}
\caption{Globular cluster colour distribution in different elliptical annuli centred on NGC\,4696. The top panels show the selected GCs, and the bottom panels show their respective background- and completeness-corrected colour distributions. The black lines in the bottom panels show a Gaussian KDE representation of the data, while the red curves show the best fit obtained from the ACS Virgo Cluster Survey \citep{pengACSVirgoCluster2006} templates transformed to $V-I$ colours.}
\label{fig:color_distributions}
\end{figure*}

\begin{figure*}
\centering
\includegraphics[width=\textwidth]{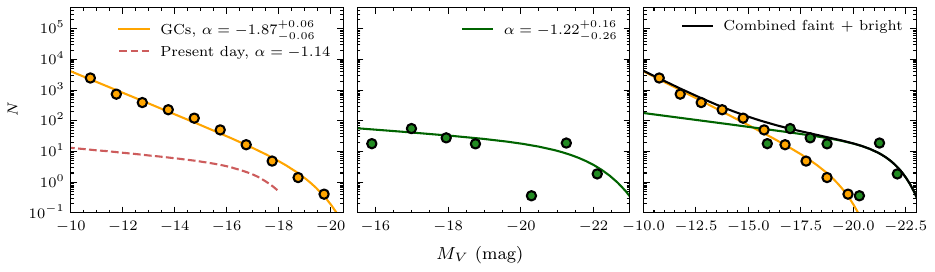}
\caption{Constraints on the evolution of the galaxy luminosity function in Centaurus. Left: reconstructed faint end of the luminosity function using the specific frequency of GCs. The red dashed line shows the present-day LF from \cite{Misgeld09}, while the orange points show the reconstructed LF from the GC specific frequency. The orange curve shows the best Schechter fit to the reconstructed LF. Middle: reconstructed bright end of the luminosity function using the colour distribution of GCs. Right: full constraint to the luminosity function, combining both methods.}
\label{fig:LF_constraint}
\end{figure*}

We repeat this analysis for the Centaurus cluster. Figure \ref{fig:color_distributions} shows the selected annuli around NGC\,4696, the corresponding GC colour distribution and the best fit obtained from the Virgo templates.
The fit in all annuli is in reasonable agreement with the data, with the highest discrepancy obtained between \qty{1.0}{}\,<\,$d$\,<\,\qty{2.0}{arcmin}.
Compared to Hydra\,I, the fitting works better in Centaurus. This is mainly due to the lack of the extended tail of very blue GCs as seen in Hydra\,I which could not be described by the Virgo templates.

The LF obtained from the template fitting is shown in the middle panel of Fig. \ref{fig:LF_constraint}.
The best-fit Schechter function has a shallow slope of $\alpha=-1.22_{-0.26}^{+0.16}$, similar to the value reported for Hydra\,I, $\alpha=-1.29_{-0.34}^{+0.21}$, although slightly better constrained.
Although they have large uncertainties, these slopes are close to the value obtained by \cite{brownAssemblyIntraclusterLight2024} for the LF of galaxies that fell into clusters after $z\sim 2$ in the Horizon-AGN simulation. The right panel of Fig. \ref{fig:LF_constraint} shows the full constraint to the past LF in Centaurus, combining the two methods discussed above.

\section{A model for the subcluster interaction}
\label{sec:model}

\begin{figure}
\centering
\includegraphics[width=0.6\columnwidth]{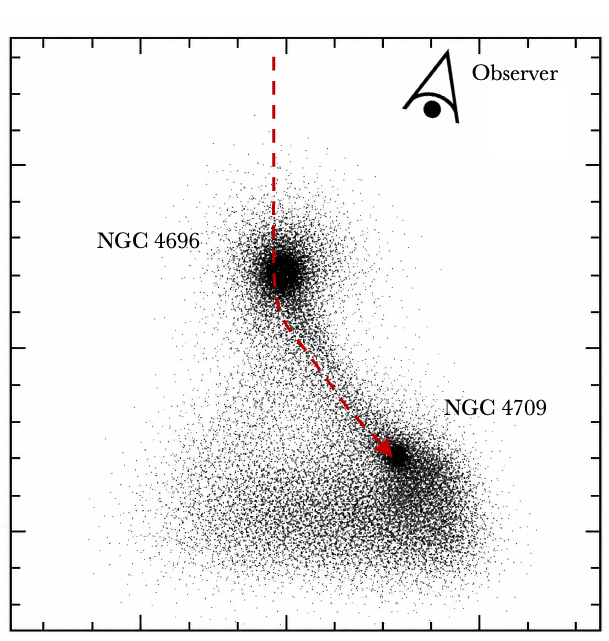}
\caption{The proposed model for the subcluster interaction, adapted from Fig. 5 from \cite{gonzalez-garciaEncountersSphericalGalaxies2005}. The trajectory followed by NGC\,4709 is sketched with a red dashed line.}
\label{fig:interaction}
\end{figure}

In the previous sections, we presented new evidence for an ongoing interaction between the two subclusters, Cen\,30 and Cen\,45, based on the GC populations in the cluster. A successful model for the interaction should be able to explain i) the large difference in line-of-sight velocity of $\Delta v_\textrm{LOS}\sim$\qty{1500}{km/s} between the two galaxies, ii) the hot X-ray-emitting gas in the region between NGC\,4696 and NGC\,4709, iii) the bridge of tidal debris (GCs, UCDs, possibly stars and dwarf galaxies) connecting the two galaxies, iv) the shallower GC density profile of NGC\,4709 compared to NGC\,4696. These features provide constraints on the geometry of the interaction.
We propose the scenario of a high-velocity, grazing hyperbolic collision between the two subclusters that occurred in the recent past, in a motion approximately along the line of sight. In this scenario, Cen\,45 passed through Cen\,30 and is now located behind it, moving away from the observer. 
The interaction is sketched in Fig. \ref{fig:interaction}.

The temperature excess in the intracluster medium between NGC\,4696 and NGC\,4709 has been interpreted as a simple shock heating of the gas due to the interaction between the two subclusters \citep{Churazov99,Walker13}.
However, bow shocks always form in front of the moving cluster where the gas compression occurs. In our scenario, the projected motion of NGC\,4709 is approximately from northeast to southwest. Its bow shock would therefore be expected to produce a region of enhanced temperature to the west of the galaxy, whereas the observed high-temperature region lies to the east.
Interestingly, when attempting to simulate the Bullet Cluster, \cite{mastropietroSimulatingBulletCluster2008} have shown that a head-on collision between two clusters produces a region of relict hot gas between the two clusters after the interaction has taken place.
This mechanism helps to explain the observed temperature excess in our proposed model for the interaction.

Figure \ref{fig:bridge_sersic_profiles} shows that the GC bridge is a real overdensity above the expected contribution from the GC systems of the two galaxies.
Numerical simulations by \cite{gonzalez-garciaEncountersSphericalGalaxies2005} show that, in a nearly head-on interaction, such a feature does not form before the pericentric passage, but rather forms after the smaller system has passed through the more massive one. In a 5:1 encounter, their simulations also suggest that the stellar particles of the more massive system are not significantly affected, which may explain the seemingly unperturbed appearance of NGC\,4696 in the optical images (see their model 5wW). 
The smaller system is more strongly affected by stripping, forming a trailing cloud of particles and puffing up its density profile.
Therefore, such an interaction may be responsible for disturbing the GC system of NGC\,4709, making it more extended, in agreement with its underdense central distribution of bGCs (Fig. \ref{fig:spatial_density_maps}) and with its shallower density profiles when compared to their counterparts in NGC\,4696 (Fig. \ref{fig:GC_profiles}).
The structure of the bridge is reminiscent of other binary cluster interactions leading to an elongated morphology of the ICL, for example NGC\,4874 and NGC\,4889 in the Coma cluster \citep{thuanPhotographicMeasurementsDiffuse1977,gerhardDetectionIntraclusterPlanetary2005,gerhardKinematicsIntraclusterPlanetary2007a}.

The low-surface-brightness features presented in Fig. \ref{fig:subtracted_images} are likely not a result of the subcluster interaction, but rather arise from processes internal to the subclusters. The geometry of the plume around NGC\,4709 resembles the accretion of a smaller system. Although we do not investigate the progenitor galaxy, one candidate is the subtracted galaxy to the southeast of NGC\,4709, which lies along the extension of the plume.
The shells around NGC\,4696 are also unlikely to have formed as a result of the subcluster interaction, but may instead have originated from a past merger or from complex reddening caused by the distribution of dust around the galaxy.

In this work, we do not attempt to constrain the extent of the GC bridge. However, if the passage occurred \qty{300}{Myr} ago at \qty{1500}{\kilo\meter\per\second}, NGC\,4709 would have travelled only \(\sim\qty{0.5}{Mpc}\) relative to NGC\,4696\footnote{We also note that, for the bridge to extend over \qty{8.5}{Mpc}, as suggested by \cite{Federle25}, NGC\,4709 would require \(\sim\qty{5.5}{Gyr}\) to cover this distance at \qty{1500}{\kilo\meter\per\second}.}. This corresponds to a distance-modulus difference of \(\lesssim 0.05\) mag, which is below the precision of current distance measurements.

Our scenario makes predictions that can be tested observationally.
First, the surplus of GCs in the bridge should be dominated by a component with a line-of-sight velocity distribution close to that of Cen\,45, as simulations suggest that the bridge originates from material stripped from the low-mass system.
Thus, the rGCs in the bridge belonging to Cen\,30 should follow the light profile of NGC\,4696, while those belonging to Cen\,45 should be more extended than the light profile of NGC\,4709.
Unfortunately, such a dataset is not available beyond the few velocity measurements of the UCDs shown in Fig. \ref{fig:subtracted_images}.
The limited number of UCD along the bridge ($\sim7$ in Cen\,30, 4 in Cen\,45) hinders any conclusion regarding the velocity structure of the bridge material, motivating spectroscopic follow-up of the GCs in this region.

Conversely, while the bulk of the stripped material from NGC\,4709 ends up in the bridge and in plumes ahead and behind the galaxy, we may expect a small population of GCs belonging to Cen\,45 around NGC\,4696. If this population exists, it is likely that it was scattered into this region as a result of the subcluster interaction (see, for example, the simulations in \citealt{bekkiDynamicalEvolutionGlobular2003}).
To date, the GC population in this region has not been studied either photometrically nor spectroscopically, although deep wide-field imaging of the region exists \citep{Federle24}.

Soon, our analysis of the Centaurus cluster with the wide-field VST data will provide further insights into the ongoing subcluster interaction (Spavone et al. in prep.).
With this dataset, we will be able to measure low-surface-brightness features and the ICL in Centaurus down to $\mu_g \sim 28$ mag/arcsec$^2$, as was done for the Hydra\,I cluster \citep{spavoneGalaxyPopulationsHydra2024a}. In particular, it will be interesting to learn how much ICL is present in the bridge region as a result of the grazing encounter. Other low-surface-brightness features may also be discovered outside our FORS coverage and will help to constrain the individual assemblies of Cen\,30 and Cen\,45.

The GC population provides a useful view of the interaction because it traces both the spatial and kinematic signatures of material stripped from the galaxies. 
The preferential contribution of bGCs to the bridge further suggests that the interaction is redistributing the metal-poor GC population into the intracluster region. 
Combined with the X-ray emission and the disturbed GC system observed in NGC\,4709, these observations support a recent interaction between Cen\,30 and Cen\,45. 
However, the three-dimensional geometry and timing of the encounter remain uncertain, and spectroscopic observations of the bridge GCs are essential to validate the proposed scenario.

% \thefontsize\small 9pt
% \thefontsize\normalsize 10pt
% \thefontsize\large 10.95pt

%-----------------------------------------------------------------

\section{Summary and Conclusions}
\label{sec:summary}

In this paper, we used deep imaging data from the FORS1 instrument at the VLT to study the past assembly history of the Centaurus cluster.
Our findings are based on the low-surface-brightness features around the two BCGs, NGC\,4696 and NGC\,4709, as well as the population of ICGCs, and can be summarised as follows:

\begin{itemize}
    \item Both BCGs show low-surface-brightness features around them: NGC\,4696 exhibits a shell system, while NGC\,4709 shows a plume emerging from the galaxy;
    \item There is a hint of a velocity structure in the UCD population between the BCGs. Those associated with Cen\,30 seem to be preferentially blueshifted with respect to NGC\,4696, while those associated with Cen\,45 seem to be redshifted with respect to NGC\,4709. 
    However, the low number statistics of the UCD sample prevents us from drawing any conclusive results.    
    \item The Voronoi-based representation of the spatial distribution of GCs shows significant substructure, most notably the bridge of GCs between NGC\,4696 and NGC\,4709 already reported in \cite{Federle25}.
    \item Red and blue GCs show different spatial distributions. Red GCs are concentrated on the two galaxies, while bGCs are more extended and dominate the intracluster space including the bridge region. The bridge geometry also appears different in the two populations, with rGCs forming a narrow filament and bGCs being more irregularly distributed.
    \item The bridge of GCs cannot be described as a simple superposition of the GC systems of the two galaxies. Comparing the rGC number density with the combined surface brightness profiles of NGC\,4696 and NGC\,4709, we still find an excess of GCs with respect to the expected population from the two galaxies.
    \item The GC number density profiles of the two galaxies also differ when plotted as a function of their respective effective radii. For both red and blue GCs, NGC\,4709 has a shallower profile than NGC\,4696, which can be interpreted as a result of the subcluster interaction occurring in the cluster.
    \item We applied the method introduced in L26 to constrain the LF of galaxies that contributed to the assembly of the cluster using the GC specific frequency and colour distribution. We find a steep faint-end slope of $\alpha=-1.87_{-0.06}^{+0.06}$, in agreement with the value reported for the Hydra\,I cluster in L26. The bright end is less constrained, with a slope of $\alpha=-1.22_{-0.26}^{+0.16}$, but is also in agreement with our previous findings.
    \item We postulate a model for the interaction between Cen\,30 and Cen\,45 that is able to explain the large line-of-sight velocity difference, the X-ray observations and the GC properties. We propose that a grazing hyperbolic encounter happened in the recent past, with Cen\,45 passing through Cen\,30 approximately along the line of sight. Numerical simulations show that such an encounter produces a bridge of debris originally associated with the smaller system, in our case NGC\,4709, as well as a cloud of stripped material that expands away from the galaxy. As a result, the GC profile of NGC\,4709 became shallower after the pericentric passage due to the stripping of many of its GCs.
    \item Our model provides testable predictions for the line-of-sight velocity of the ICGCs. If the GC bridge indeed originated from NGC\,4709, its velocities should be compatible with that of Cen\,45. Also, A small population of Cen\,45 GCs may have been scattered around NGC 4696 as a result of the subcluster interaction.
\end{itemize}

These predictions can be tested quite simply with spectroscopic follow-up of the GCs in the bridge and those surrounding NGC\,4696 and NGC\,4709.
Wide-field and high-multiplex spectrographs such as the upcoming MOONS instrument at the VLT present promising means to obtain these data, ultimately confirming or ruling out our proposed model for the interaction in the Centaurus cluster.

\begin{acknowledgements}
      Based on observations collected at the European Organisation for Astronomical Research in the Southern Hemisphere under ESO programme 67.A-0358(A).
      FSL thanks the support from the IMPRS on Astrophysics at the LMU Munich. SF gratefully acknowledge support by ANID through Fondecyt Regular 1240755. MC acknowledges support from ASI–INAF grant no. 2024-10-HH.0 (WP8420), the ESO Scientific Visitor Programme, and INAF GO-grant no. 12/2024.
\end{acknowledgements}

\bibliographystyle{aa} % style aa.bst
\bibliography{zoterobib} % your references Yourfile.bib

\begin{appendix} %First appendix

\section{Point source selection and completeness correction}
\label{app:point_source_selection}

At the distance of the Centaurus cluster, GCs are observed as unresolved point sources.
The morphological parameters obtained from the \textsc{daophot} PSF photometry, sharpness and $\chi$, can be used to select point sources.
Sharpness is a parameter that quantifies the difference between the square of the width of an object and the square of the width of the PSF, with single stars having sharpness close to zero and extended objects having larger absolute values;
$\chi$ corresponds to the goodness of the PSF fit, quantifying the pixel-to-pixel scatter from the fitting residuals.
% Point sources have sharpness close to zero.
We therefore restrict our sources to small absolute values of sharpness, as well as low values of $\chi$.
Finally, the photometric error is also constrained to values smaller than \qty{0.2}{mag}.

\begin{figure}
    \centering
    \includegraphics[width=\columnwidth]{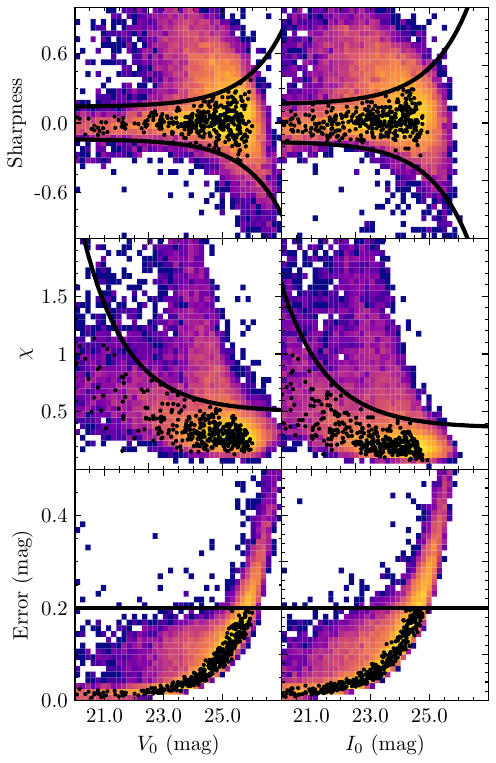}
    \caption{Density distribution of the morphological parameters, sharpness and $\chi$, and of the photometric error as a function of magnitude, for all detected sources in our images. 
    Brighter colours represent higher densities of points.
    The adopted cuts in each parameter are represented as solid black lines, while the black points show a subsample of the selected point sources.}
    \label{fig:point_source_selection}
\end{figure}

Figure \ref{fig:point_source_selection} shows the adopted cuts in sharpness, $\chi$ and photometric error in the $V$ and $I$ bands, represented by the thick solid lines.
Accepted sources are those that satisfy all the criteria in both bands simultaneously.
A small subset of selected sources are plotted as black points, to help visualise their location with respect to the full sample of detected sources.
The final sample of point sources comprises 27021 objects.

The photometric completeness was quantified from completeness curves obtained by performing injection-recovery experiments of artificial stars.
The galaxy-subtracted images were injected with 100 artificial stars, followed by the same PSF photometry performed on the original images.
The artificial stars follow a uniform distribution in magnitude and in space, and each injection was repeated 100 times. Then, we analyse the fraction of recovered stars in each magnitude bin.

\begin{figure}
\centering
\includegraphics[width=\columnwidth]{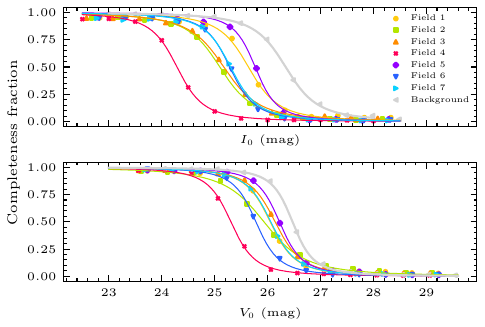}
\caption{Completeness fractions recovered from artificial star experiments for all fields in the $V$ and $I$ bands. }
\label{fig:completeness}
\end{figure}

Figure \ref{fig:completeness} shows the completeness curves for all fields in $V$ and $I$.
The symbols show the recovered fraction in each magnitude bin, while the solid lines show the best fit for a Pritchet function \citep{mclaughlinSpatialStructureM871994}.
The completeness levels are similar between all fields except for field 4.
From these fits, we obtain the 95\% completeness magnitude in each band, $V_\textrm{95}$ and $I_\textrm{95}$, which are used to obtain the GC number density correction factors.

\begin{figure*}
\centering
\includegraphics[width=\textwidth]{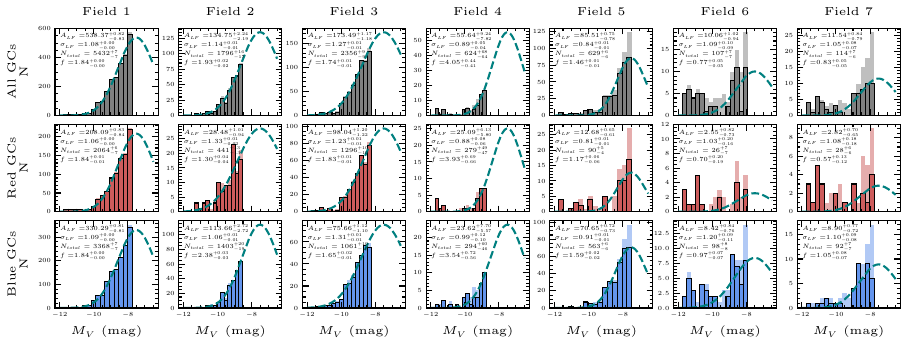}
\caption{Gaussian fits to the background-subtracted GC luminosity function in each field for all (top row), red (middle row) and blue (bottom row) GCs. In each panel, the lighter and darker histograms represent the observed and the background subtracted samples, respectively. For each fit, we annotate the normalisation and dispersion of the Gaussian, $A_{LF}$ and $\sigma_{LF}$, the total number of GCs expected in the field, $N_{total}$, and the resulting correction factor, $f$.} 
\label{fig:GCLF_fits}
\end{figure*}

The correction factors were obtained from the GC luminosity function (GCLF) in each field. As explained in the main text, we determine a limiting $V$-band magnitude where the completeness is at least 95\% in both bands. Namely, $V_\textrm{lim} = \textrm{min}(V_\textrm{95}, I_\textrm{95} + 0.8)$, where \qty{0.8}{mag} is an adopted lower limit on the colour of our GC sample.
Then, we subtract the number of background objects in each magnitude bin from the GC luminosity function up to $V_\textrm{lim}$, and fit a Gaussian function centred in $M_V=$ \qty{-7.5}{mag} and with a Gaussian $\sigma$ free to vary between 0.80 and \qty{1.35}{mag}.
The total number of GCs expected inside the field corresponds to the integral of the fitted Gaussian divided by the magnitude bin size that was used for the fitted luminosity function.
Finally, the correction factor is obtained as the ratio between this total number of GCs and our observed sample size in each field.
Figure \ref{fig:GCLF_fits} shows the Gaussian fits to the GCLF in each field, as well as the resulting correction factors indicated in the legend.

% \begin{sidewaysfigure*}
%     \centering
%     \includegraphics[width=\textwidth]{figs/appendix_gclf_fits.pdf}
%     \caption{Gaussian fits to the background-subtracted GC luminosity function in each field for all (top row), red (middle row) and blue (bottom row) GCs. In each panel, the lighter and darker histograms represent the observed and the background subtracted samples, respectively. For each fit, we annotate the normalisation and dispersion of the Gaussian, $A_{LF}$ and $\sigma_{LF}$, the total number of GCs expected in the field, $N_{total}$, and the resulting correcting factor, $f$.} 
%     \label{fig:GCLF_fits}
% \end{sidewaysfigure*}

\end{appendix}

\end{document}